\documentstyle[epsfig,prl,aps,twocolumn]{revtex}
\begin{document}

\draft
\title{\bf An attempt to incorporate separate chemical and thermal
  freeze-outs in hydrodynamics}
\author{N. Arbex$^a$, F. Grassi$^a$, Y.Hama$^a$ and O. Socolowski Jr.$^b$}
\address{$^a$ Instituto de F\'{\i}sica, Universidade de S\~{a}o Paulo, \\
C. P. 66318, 05315-970 S\~{a}o Paulo-SP, Brazil\\
$^b$ Subatech - Ecole des Mines de Nantes,\\
4, rue Alfred Kastler - La Chantrerie - BP 20722
44307 Nantes cedex 3, France}
\date{\today} 
\maketitle

\begin{abstract} 
From an analysis of  various types of data  obtained in relativistic
nuclear  collisions,  
the following  picture
has emerged in thermal  and  hydrodynamical descriptions:  
as the fluid expands and
cools,  particles first undergo  a chemical freeze-out at 
$T_{ch.f.}\sim 160-200$ MeV then  a thermal freeze-out at  
$T_{th.f.}\sim 100-140$ MeV. In this paper we show how to incorporate
these  separate freeze-outs consistently in a  hydrodynamical code
via a modified equation of state (general case) or via a modified
Cooper-Frye formula (particular case of $T_{ch.f.}$  close to
$T_{th.f.}$ or few particle species undergoing early chemical freeze-out). 
The modified equation of state  causes faster cooling and
may
have sizable impact on the predicted values  of observables.
\end{abstract}


\section{Introduction}

The behavior of strongly interacting matter under extreme conditions
of pressure and temperature is the subject of the research programs at
CERN (SPS) and Brookhaven (AGS and RHIC).
Hydrodynamical and thermal models have been used extensively to describe
data from these collisions and the following picture has emerged
(see e.g. \cite{shQM99,heQM99}) from  a study  of collisions  at SIS,
AGS,  SPS energies,  with a  variety of targets  and projectiles
(for a compilation see  e.g. \cite{cl99}). It is also expected to
hold at RHIC and LHC energies.
The dense and hot fluid expands and cools until chemical freeze-out
occurs for some species of particles. Namely these particles stop
having inelastic collisions so that their abundances are frozen.
Therefore by studying the abundances of these chemically frozen
particle species, one learn the conditions at chemical freeze-out.
 For example, at CERN, $T_{ch.f.} \sim 160-200$ MeV \cite{sollfranksqm97}.
The fluid goes on cooling until thermal freeze-out happens.
Precisely, particles stop having elastic interactions and so the shape of
their momentum distribution is fixed. Therefore by studying
these spectra, one gets information about the conditions at thermal
freeze-out. For example at CERN, using various types of particles 
\cite{xu96} or
combining information about the spectrum of a single particle species
and its Bose-Einstein correlations \cite{wi98,wi99}, one extracts 
$T_{th.f.}\sim 100-140$ MeV. There are some deviations to this picture.
For example,
some particles like the $\Omega$
may  undergo both freeze-outs almost together and early 
(due to their small cross section).

Though this picture
is simple and consistent with data, its
theoretical justification  needs further scrutiny.
First, in a way, this picture works too well as a statistical description
 seems to apply even to very
elementary systems \cite{becattini}. This has 
been debated a lot yet remains an open question.
Second, in this description, it is assumed that particles make {\em
  sudden} freeze-outs. For example, when they cross the 180 MeV
temperature
three-dimensional surface in the fluid they  immediately stop interacting
inelastically. In reality one expects  that this should
happen over a
certain length. A formalism to account for finite  freeze-out volumes
 and  the subsequent new data interpretation have been presented
 in \cite{grassi}.
Third, 
in the freeze-out scenario,  when computing particle
distribution with the usual Cooper-Frye formula \cite{CF}, there may
be particles contributing negatively, corresponding to particles that
are in the frozen out region and re-entering the interacting region.
Physically this should not happen but in the calculation, it may.
Ways to deal with this problem and how it affects particle data
interpretation can be found in
 \cite{csernai}.

Leaving aside these problems for further studies,
the objective of this paper is the following. In the past,
cascade event generators (Fritiof, Venus, RQMD, ARC, etc)
were employed in particular by experimental groups, to study
 their data. 
Recently, simple thermal and hydrodynamics-inspired models 
have been used increasingly. It seems useful therefore
to start developing more sophisticated hydrodynamical codes \cite{sh98} 
to extract physical information from data.
Though data call for separate chemical and thermal freeze outs, no 
hydrodynamical code so far includes chemical and thermal freeze outs
self-consistently. Namely, the effect of the early chemical freeze out on the
fluid expansion is never taken into account.
In this paper,
we discuss how to incorporate separate chemical
and thermal freeze-outs in a hydrodynamical code and show that in
certain cases, this will have
sizable impact on  the predicted values of observables.

 Finally let us mention that 
early universe and relativistic heavy ion 
collisions call for different treatments for the following reason.
In the early universe,
 the expansion rate ($H \sim 10^4 s^{-1}$)  at the QCD phase transition is
many orders of magnitude smaller than for
 relativistic heavy ion  collisions ($H \sim 10^{21}-10^{23} s^{-1}$).
In both cases, a typical reaction rate for hadrons
is $\Gamma = \sigma n v_{rel}
\sim 1 fm^{-1}= 10^{23} s^{-1}$. Therefore while  in the early
universe, chemical and thermal equilibrium betweeen hadrons must have prevailed
($\Gamma >> H$), this  is not the case for relativistic
heavy ion collisions ($\Gamma$ may be   $\sim H$). 
Moreover in relativistic heavy ion collisions, we expect separate
 chemical and thermal freeze-outs
 (i.e. at distinct temperatures) because
 inelastic collisions (responsible for chemical equilibrium) need
 higher center of mass energy
to be operative in  general, compared to elastic collisions
(responsible for thermal equilibrium). This distinction needs not be
made in the early universe (both rates $\Gamma_{inel}$ and $\Gamma_{el}$
being much larger than the expansion rate).

\section{Inclusion of separate chemical and thermal freeze-outs in the
 hydrodynamical equations}

To simplify the discussion, we use a known and simple hydrodynamical
model, Bjorken one-dimensional boost invariant model\cite{bj} . 
Before chemical freeze-out, the fluid evolution is governed by the
hydrodynamical equations
\begin{eqnarray}
\frac{\partial \epsilon}{\partial t}+\frac{\epsilon+p}{t} & = & 0\\
\frac{\partial n_B}{\partial t} + \frac{n_B}{t} & = & 0.
\end{eqnarray}
The last equation can be solved easily
\begin{equation}
n_B(t)=\frac{n_B(t_0) t_0}{t}.
\end{equation}
The first equation
 must be completed by the choice of an equation of
state $p(n_B,\epsilon)$, for the pressure as function of the net baryon
 density
and energy density.

Also to simplify the discussion, we suppose that chemical or thermal
freeze-out occurs at some fixed temperature (as often assumed in
the analysis of experimental data). 
Attempts to incorporate more physical freeze-out conditions
have been carried out \cite{leh89,na92,he87,le88,sh98}
and in principle might be incorporated in the
scheme described below.

When the fluid temperature has decreased to some temperature $T_{ch.f.}$,
(which corresponds to  a certain time $t_{ch.f.}$), some particle species get
their abundances frozen. To fix ideas, we suppose that $\Lambda$ and
$\bar{\Lambda}$ are in this situation. Then in addition to the above
hydrodynamical equations, we introduce separate conservation laws for
these two types of particles for time $t>t_{ch.f.}$, namely
\begin{eqnarray}
\frac{\partial n_{\Lambda}}{\partial t} + \frac{n_\Lambda}{t} & = & 0\\
\frac{\partial n_{\bar{\Lambda}}}{\partial t} + \frac{n_{\bar{\Lambda}}}{t} & = & 0.
\end{eqnarray}
These equations have solutions of the same form as (3) but with $t_0$
 substituted by $t_{ch}$.
Therefore what remains to be done is to solve the energy-momentum
equation (1) with a {\em modified equation of state}, to account for the
particles who make an early chemical freeze-out.

We suppose that the fluid is a gas of non-interacting resonances.
Then for particle species $i$, using an expansion in term of modified Bessel
functions \cite{landau} to allow the study of  
their limit more easily in (9-11),
\begin{eqnarray}
n_i & = & \frac{g_i m_i^2 T}{2 \pi^2} \sum_{n=1}^{\infty} (\mp)^{n+1}
\frac{e^{n \mu_i/T}}{n} K_2(n m_i/T)\\
\epsilon_i & = & \frac{g_i m_i^2 T^2}{2 \pi^2} \sum_{n=1}^{\infty} (\mp)^{n+1}
\frac{e^{n \mu_i/T}}{n^2}[3 K_2(n m_i/T)\\
 & + & \frac{n m_i}{T} K_1(n m_i/T)]
\nonumber
\\
p_i  & = & \frac{g_i m_i^2 T^2}{2 \pi^2} \sum_{n=1}^{\infty} (\mp)^{n+1}
\frac{e^{n \mu_i/T}}{n^2} K_2(n m_i/T)
\end{eqnarray}
where $m_i$ is the particle mass, $g_i$, its degeneracy and
$\mu_i$, its chemical potential, the minus sign holds for fermions and
plus for bosons. In principle each particle
species  $i$ making early chemical freeze-out has a chemical potential
associated to it; this potential controls the conservation of the
number of particles of type $i$. For particle species not making early
chemical freeze-out, the chemical potential is of the usual type,
$\mu_i=B_i \mu_B +S_i \mu_S$, where $ \mu_B$ ($\mu_S$) ensures the
conservation of baryon number (strangeness) and $B_i$ ($S_i$) is the baryon
(strangeness)
number of particle of type $i$. 
So the modified equation of state depends not only on $T$ and $\mu_B$ but also
$\mu_{\Lambda}$, $\mu_{\bar{\Lambda}}, etc $. 
(the notation ``$etc$''  stands for all the other particles making early
chemical freeze-out) \cite{bebie}. This complicates  the
 hydrodynamical problem, however we can note the following.

If $m_i-\mu_i>> T$ (the density of type $i$ particle is low)
and  $m_i >> T$, (these relations should hold for
all particles except pions and we checked them  for various times and
particle types)
the
following
approximations can be used
\begin{eqnarray}
n_i & = & \frac{g_i}{2 \pi^2} \sqrt{\frac{\pi}{2}} (m_i T)^{3/2}
e^{(\mu_i-m_i)/T}
\left(1+\frac{15 T}{8 m_i }
                  + \frac{ 105 T^2}{128 m_i^2}+...
\right)
\\
\epsilon_i & = & n_i m_i
\left(1+\frac{3 T}{2 m_i }
                  + \frac{ 15  T^2}{8 m_i^2}+...
\right)
\\
p_i  & = & n_i T
\end{eqnarray}
We note that $\epsilon_i$ and $p_i$ are written in term of $n_i$ and T.
Therefore we can work
 with the variables $T, \mu_B,n_{\Lambda},n_{\bar{\Lambda}},etc$,
rather than $T,\mu_B,\mu_{\Lambda},\mu_{\bar{\Lambda}}, etc$.
The time dependence of $n_{\Lambda},n_{\bar{\Lambda}}, etc$ is known 
as discussed already.
So the modified equation of state can be computed from $t$, $T$ and 
$\mu_B$. 

The scheme presented above can easily be generalized to particles making 
chemical freeze-out at different times (using different $t_{ch.f.}$)
and particles doing 
chemical and thermal freeze-outs together (whose contribution drop out
of the equation  of state). 
One can show that entropy is conserved even in the presence of an early
chemical freeze-out when the hydrodynamical equations are satisfied by
a perfect fluid \cite{bebie}.

For illustration,
we present results using in the equation of state, the basic
multiplets of resonances (pseudoscalar meson octet plus singlet,
vector meson octet plus singlet, baryon octet and baryon decuplet)
and supposing that the early chemical
freeze-out occurs
at 180 MeV, a value typical for SPS energy more or less 
independently of the projectile \cite{cl99}.
We use for initial conditions $T_0=\mu_{B\,0}=200$ MeV, $\tau_0=1$ fm,
so that $\mu_{B\,ch.f.}=210$ MeV, which is in agreement with results
 for S or Pb at SPS  \cite{cl99}.
 In figure 1,
we compare the behavior of $T$ and $\mu_B$ as function of $t$, 
obtained from the
hydrodynamical equations using the  modified equation of
state and the unmodified one. 
For the modified equation of state,
we considered two scenarios: (I) all
strange particle in the basic multiplets, (II)   all strange particles
except $K$ and $K^*$'s, make an early chemical freeze-out.
This is a conservative estimate, it is possible for example that the
pions make an early freeze-out \cite{sh98}.
Comparing scenarios I and II, we note that 
if more particles undergo early chemical freeze-out, stronger
effects for $T(t)$ and $\mu_B(t)$ are seen.
We concentrate on I thereon.
 We see that the deviations between  I and the 
unmodified equation of state case increase with time.
In particular from this figure, 
if the thermal freeze-out occurs at 110 MeV, the 
thermal freeze-out time is 13 fm for the modified equation of state and
20 fm for the unmodified one; the corresponding baryonic potentials
are not very different, 405 and 375 MeV respectively. 
If the thermal freeze-out
occurs earlier, say at 140 MeV, the difference in the thermal freeze-out
 times would be much less. An immediate consequence of this, is that
the thermal freeze-out volume (in our case simply proportional to time)
may be much smaller for the modified equation of state.
We expect
 this conclusion to hold qualitatively
even in the presence of transverse
expansion: in this case, expansion is faster and the thermal freeze-out
 temperature is reached faster, so the thermal freeze-out times for the
modified and unmodified equation of state are less different, however
there is a competing effect for the thermal freeze-out volumes,
they now scale with higher powers of time.
We also expect this conclusion to hold at RHIC energies, with $T_{ch}$
still of order 180 MeV (it cannot be much higher since a transition
to quark-gluon plasma is expected at about this temperature from
lattice gauge simulations) but a lower value of 
$\mu_{B,ch}$ (less
baryon stopping is expected at RHIC than SPS), 
the value of $T_{th}$ may be a little
smaller than at SPS \cite{ha91,na92} and  
$\mu_{B,th}$ will be  higher than $\mu_{B,ch}$ (cf. figure 1).
{\em We conclude that
 if the chemical and thermal freeze-out
temperatures are very different} (in our simplified case, 180 and 110
MeV) {\em or if many particle species make an early chemical freeze out}
(e.g. also pions),
{\em it is important
to take into account the effect of the
early chemical freeze-out on the equation of state} to make predictions
for observables which depend on thermal freeze-out volumes, for example
 particle abundances and  eventually particle correlations.
This is our main result.
\begin{figure}[htbp]
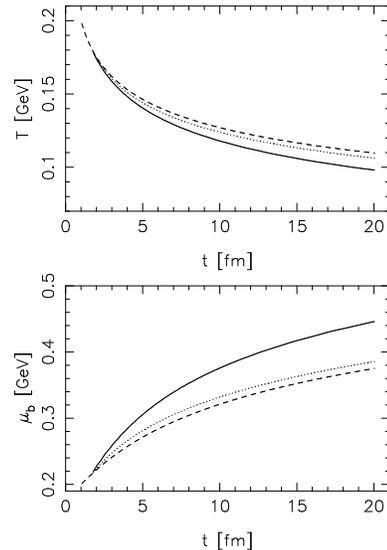

\begin{center}
\epsfig{file=T.ps,height=5.cm,angle=-90}\hfill
\epsfig{file=ub.ps,height=5.cm,angle=-90}
\caption{\footnotesize  $\mu_B$ and T as function of
  time in the case where all particles
have simultaneous freeze-outs (dashed line) and (I) all strange particles
in basic multiplets
 make an early chemical freeze-out (continuous line), (II)
 all strange particles except $K$ and $K^*$'s
 make an early chemical freeze-out (dotted line).}
\end{center}
\end{figure}

\section{Particular case where $T_{ch}\sim T_{th}$ or few particle species
undergo early chemical freeze out}

If the 
chemical and thermal freeze-out
temperatures are not very different (say 180 and 140 MeV), or if few
particle
species make the early freeze out,
one can proceed as follows.
One can use an unmodified equation of state in a
hydrodynamical code and to account for early chemical freeze-out of species
$i$, when the number of type $i$ particles was fixed,
use a modified Cooper-Frye formula
\begin{equation}
\frac{E d^3N_{i}}{dp^3}= 
\frac{N_{i}(T_{ch.f.})}{N_{i}(T_{th.f.})} \times
\int_{S_{th.f.}} d\sigma_{\mu} p^{\mu} f(x,p).
\end{equation}
The second factor on the right hand side is the usual one and it
gives the shape of the spectrum at thermal freeze-out, the first factor
is a normalizing term introduced such that upon
integration on momentum $p$, the number of particles of type $i$ is $N_i(T_{ch.f})$.
For illustration, we show results obtained with 
the hydrodynamical model, HYLANDER-PLUS 
\cite{nel}. 
It provides a numerical solution of the relativistic hydrodynamical
equations
in (3+1) dimensions with axial symmetry (for details, see
\cite{nel,hyd,udo1,sch1}). It gives a good description of
single-particle-rapidity 
data, transverse-momentum spectra of $h^-$, $\pi^-$, $p$, $\bar{p}$,
$K^0$, 
 $\pi^-/\pi^+$ and pion correlation data at CERN energies, 
 for an appropriate choice of the initial 
conditions,
 an equation of state incorporating a first order phase transition
 and a freeze-out temperature of 139 MeV.
In figures 2 and 3, we show results obtained for 
$\Lambda$, $\bar{\Lambda}$  and  
$\Xi$, $\bar{\Xi}$  respectively (neglecting resonance decays) and
data\cite{data}.

\begin{figure}[hbtp]
\begin{center}
\epsfig{file=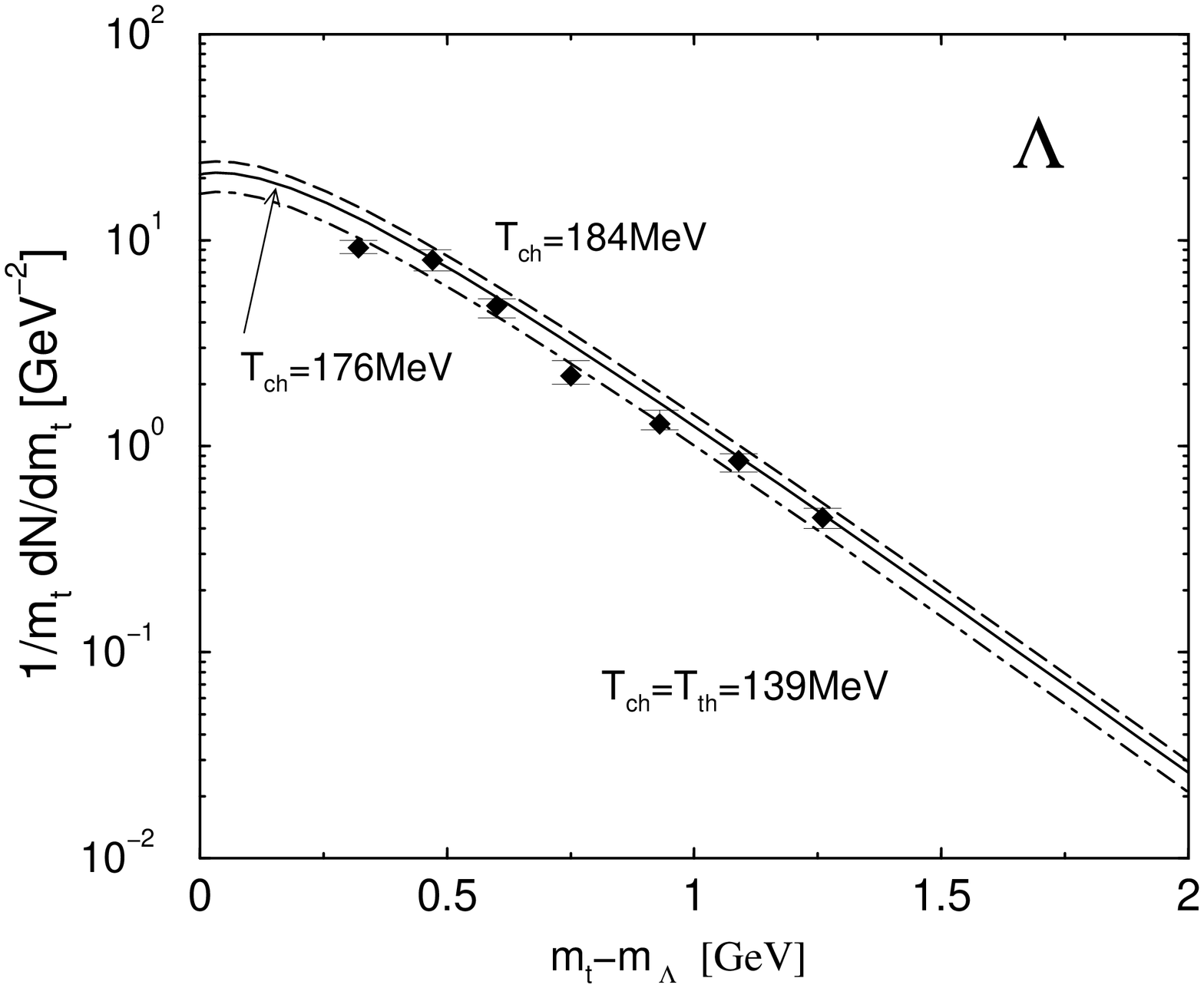,height=5.cm,angle=-90}\hfill
\epsfig{file=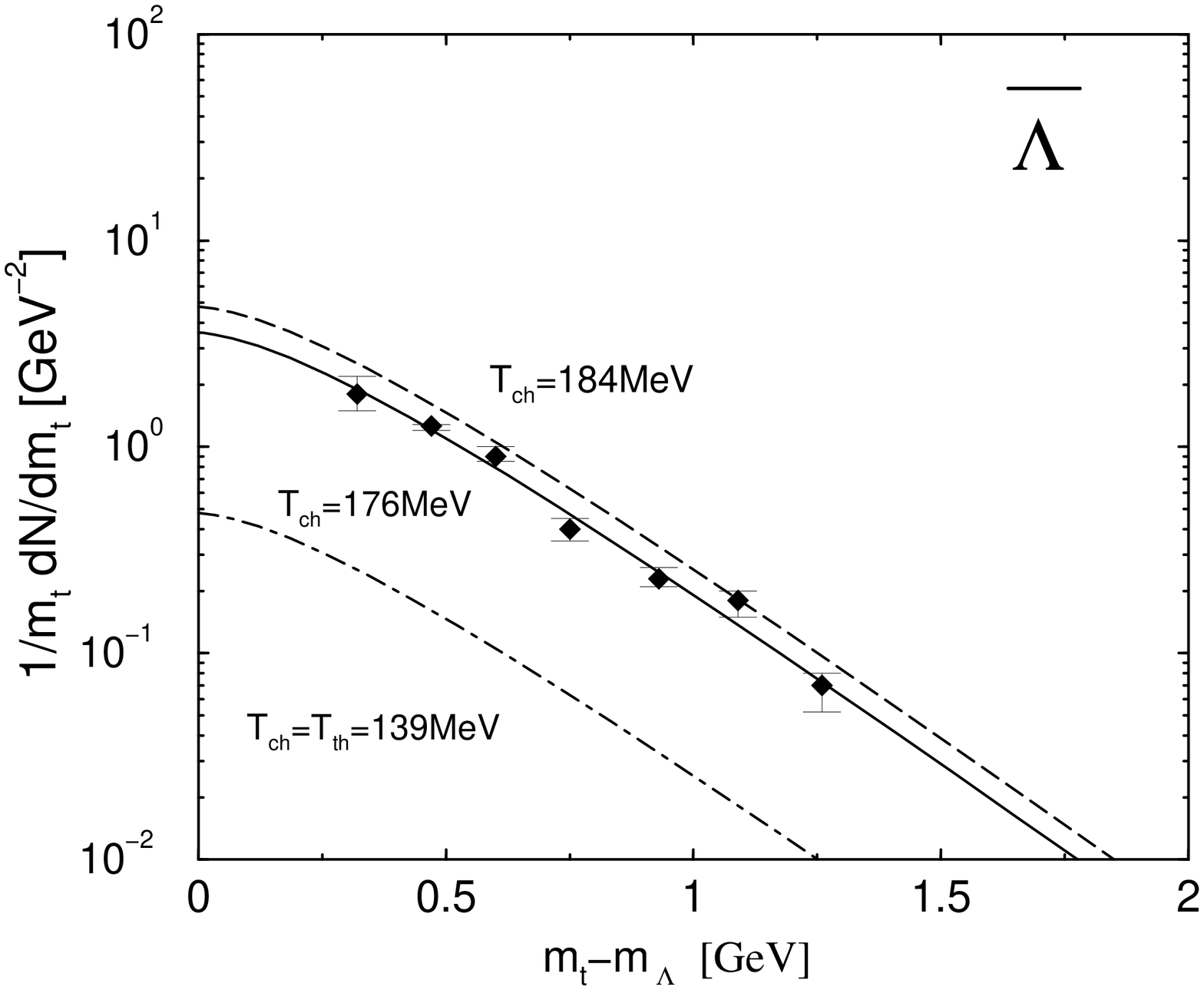,height=5.cm,angle=-90}
\caption{\footnotesize   Results from HYLANDER-PLUS
for the transverse momentum
spectra of $\Lambda$  and $\bar{\Lambda}$ compared with data [22]
for simultaneous freeze-outs at $T_{ch.f.}=T_{th.f.}=139$
MeV  (dash-dotted line)  as well  as separate  freeze-outs at
$T_{ch.f.}=176$ MeV (continuous  line) or  $T_{ch.f.}=184$ MeV
(dashed line) and  $T_{th.f.}=139$ MeV.}
\end{center}
\end{figure}

\begin{figure}[htbp]
\begin{center}
\epsfig{file=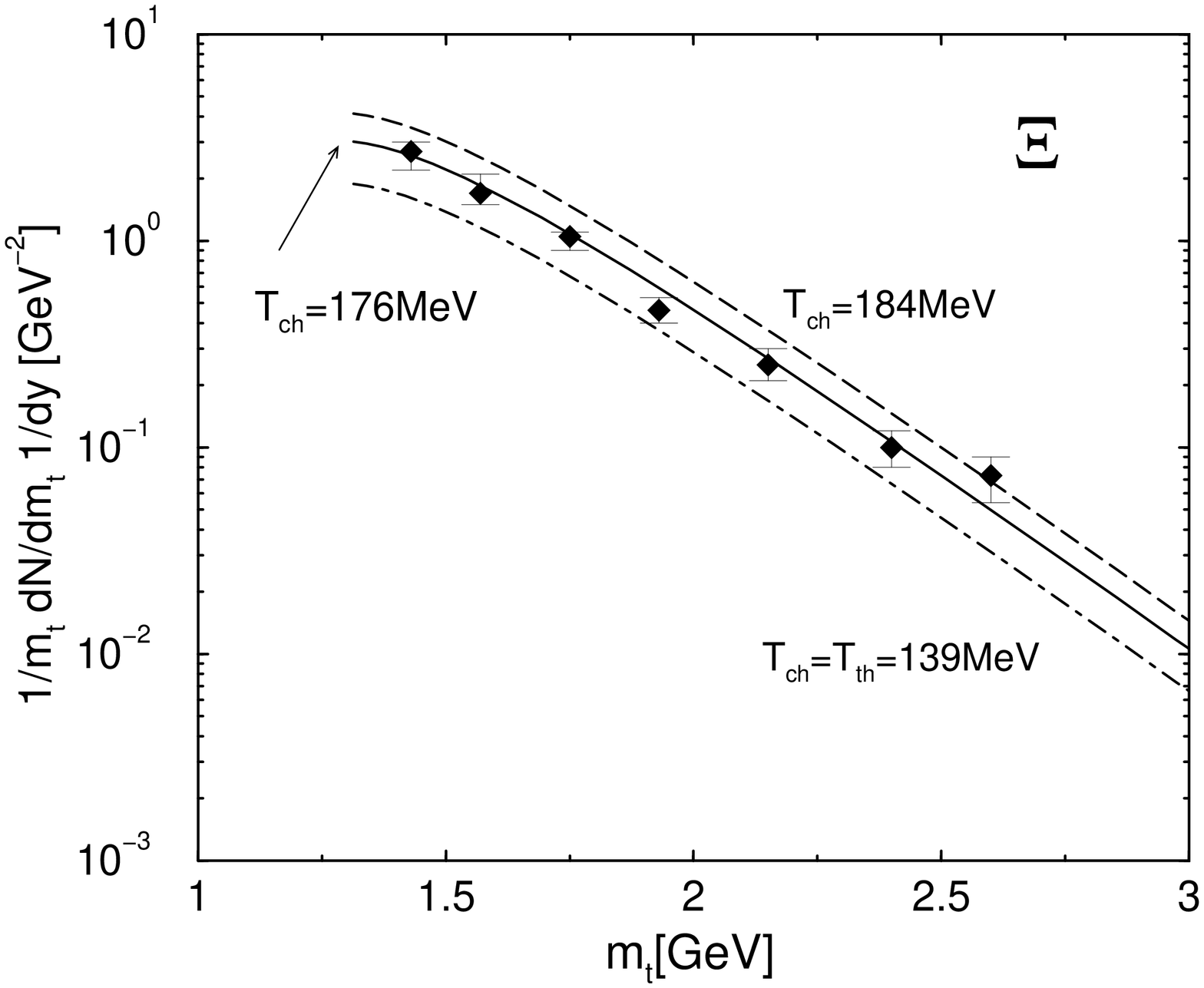,height=5.cm,angle=-90}\hfill
\epsfig{file=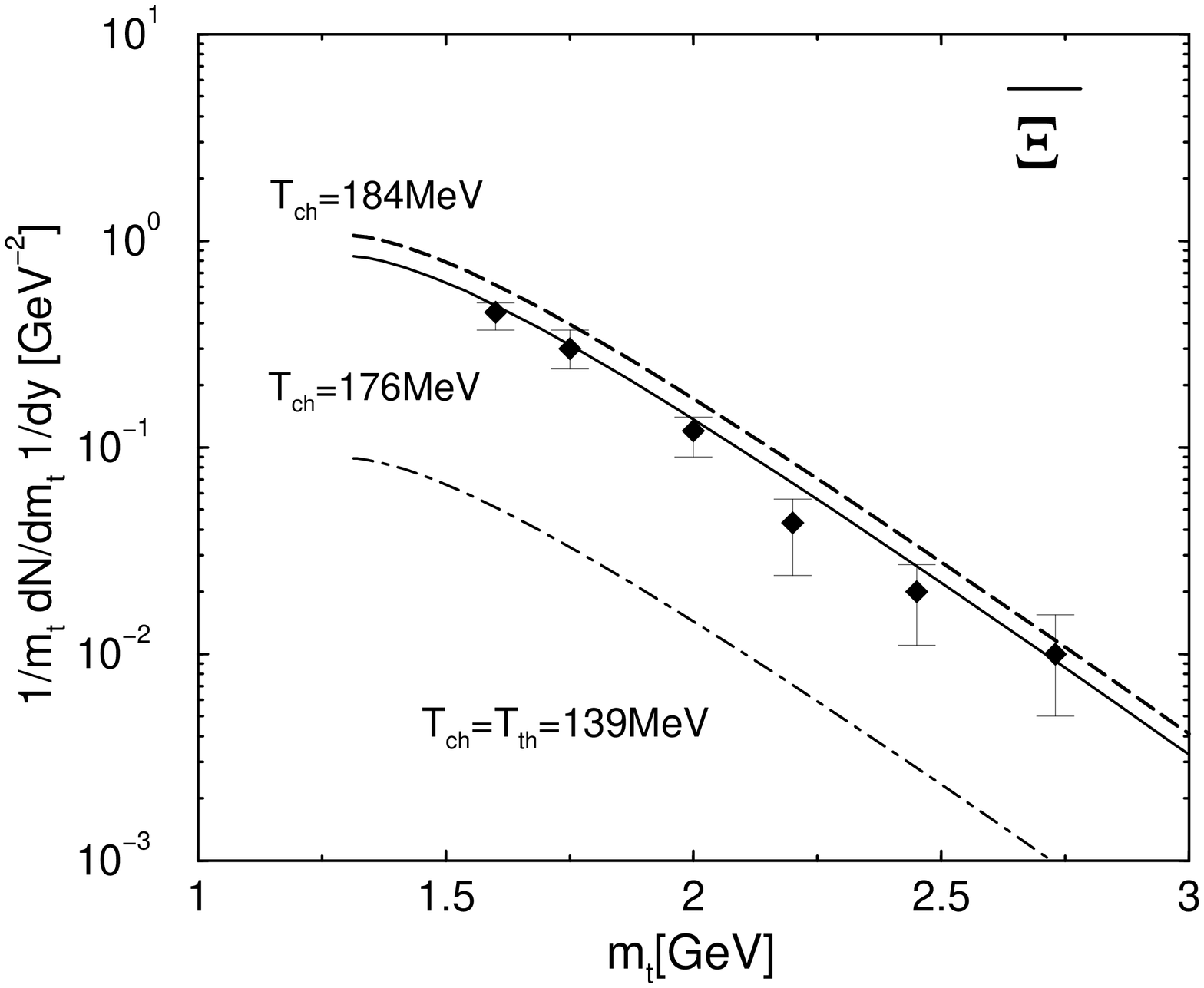,height=5.cm,angle=-90}
\caption{\footnotesize  Same  as figure 2  but  for 
$\Xi$ and $\bar{\Xi}$.}
\end{center}
\end{figure}

\noindent We see that both their shapes and
abundances can be reproduced for $T_{ch.f.}=176$ MeV and
 $T_{th.f.}=139$ MeV, while simultaneous freeze-outs at 
$T_{ch.f.}=T_{th.f.}=139$ MeV would yield
the correct shapes but too few particles.
Therefore,results with HYLANDER-PLUS and a modified Cooper-Frye
formula (12) support  the
separate freeze-outs picture. However in this code,
$T_{th.f.}$ is fixed to  139 MeV while as already
mentioned, some data seem to imply lower  thermal freeze-out temperatures. 
In the next generation of hydrodynamical codes, it is desirable to
consider  a wider range  of  $T_{th.f.}$.

\section{Conclusion}

In summary,  we showed how to incorporate separate chemical and
thermal freeze-outs in a hydrodynamical code
via a modified equation of state in section II (general case) or via a modified
Cooper-Frye formula in section III (particular case of $T_{ch.f.}$  close to
$T_{th.f.}$  or few particle species undergoing early chemical freeze-out). 
The modified equation of state  causes faster cooling and
may
have sizable impact on the predicted values  of observables.

This work was partially supported by FAPESP (2000/04422-7,2000/05769-0)
and CNPq (proc. 300054/92-0).

\end{document}